\documentclass[aps,prl,twocolumn,superscriptaddress,groupedaddress,amsmath,amssymb]{revtex4}
\usepackage{graphicx}
\usepackage{amsmath}
\usepackage{xcolor}

\renewcommand{\eqref}[1]{Eq.~(\ref{#1})}

\begin{document}
\title{Photon State Tomography for Two-Mode Correlated Itinerant Microwave Fields}
\author{C.~Eichler}
\affiliation{Department of Physics, ETH Z\"urich, CH-8093, Z\"urich, Switzerland.}
\author{D.~Bozyigit}
\affiliation{Department of Physics, ETH Z\"urich, CH-8093, Z\"urich, Switzerland.}
\author{C.~Lang}
\affiliation{Department of Physics, ETH Z\"urich, CH-8093, Z\"urich, Switzerland.}
\author{M.~Baur}
\affiliation{Department of Physics, ETH Z\"urich, CH-8093, Z\"urich, Switzerland.}
\author{L.~Steffen}
\affiliation{Department of Physics, ETH Z\"urich, CH-8093, Z\"urich, Switzerland.}
\author{J.~M.~Fink}
\affiliation{Department of Physics, ETH Z\"urich, CH-8093, Z\"urich, Switzerland.}
\author{S.~Filipp}
\affiliation{Department of Physics, ETH Z\"urich, CH-8093, Z\"urich, Switzerland.}
\author{A.~Wallraff}
\affiliation{Department of Physics, ETH Z\"urich, CH-8093, Z\"urich, Switzerland.}
\date{\today}
\begin{abstract}
Continuous variable entanglement between two modes of a radiation field is usually studied at optical frequencies. As an important step towards the observation of entanglement between propagating microwave photons we demonstrate the experimental state reconstruction of two field modes in the microwave domain.
In particular, we generate two-mode correlated states with a Josephson parametric amplifier and detect all four quadrature components simultaneously in a two-channel heterodyne setup using amplitude detectors. Analyzing two-dimensional phase space histograms for all possible pairs of quadratures allows us to determine the full covariance matrix and reconstruct the four-dimensional Wigner function. We demonstrate strong correlations between the quadrature amplitude noise in the two modes. Under ideal conditions two-mode squeezing below the standard quantum limit should be observable in future experiments.
\end{abstract}
\maketitle
State tomography for more than a single mode of a radiation field allows to characterize photon sources that display entanglement between propagating photons. At optical frequencies well-established multi-mode state reconstruction techniques based on single photon counters exist. These have already allowed for a variety of experiments that demonstrated entanglement and the EPR paradox for continuous  variables states \cite{Einstein1935, Reid1988, Ou1992, Babichev2004,Vasilyev2000}. A well-known representative for the class of optical entangled states is the two-mode squeezed vacuum \cite{Loudon1987} which has been used as a resource for
continuous variable quantum computation, cryptography and teleportation experiments \cite{Braunstein2005,Grosshans2003,Furusawa1998}.

In the microwave domain, state tomography experiments have recently been realized  for single itinerant field modes \cite{Mallet2010,Eichler2010,Menzel2010}. In addition,  multi-mode state reconstruction for intra-cavity fields of two spatially separated cavities has been demonstrated \cite{Wang2010}. Here we present heterodyne state tomography techniques that allow for the state reconstruction of \emph{two itinerant} modes. We apply these techniques to reconstruct the Wigner function of a two-mode correlated state generated in a Josephson parametric amplifier.

The general interest in parametric amplification \cite{Yurke1984,Caves1982} has steadily grown in the recent past \cite{Castellanos2007,Bergeal2010,Tholen2009} because it allows to investigate the quantum properties of microwave radiation \cite{Mallet2009}, superconducting qubits \cite{Vijay2010} and nanomechanical oscillators \cite{teufel2009} at high signal-to-noise-ratios. Noise squeezing at microwave frequencies  has been demonstrated for these amplifiers in the degenerate  \cite{Castellanos2007} as well as in the non-degenerate case of parametric amplification \cite{Bergeal2010a}.

In this letter we present state tomography measurements for two output field modes of a parametric amplifier. The two modes emitted from a single broadband resonator mode are separated in frequency space \cite{Yonezawa2007,Kamal2009}. Complementary, one can analyze squeezing correlations between two spatially separated modes by sending two squeezed states trough a balanced beamsplitter \cite{Mallet2010} or by using two parametrically coupled resonators with different output ports \cite{Bergeal2010a}.
Here, we first  describe the measurement setup,  discuss the device parameters, and characterize the system as a phase insensitive amplifier. We then measure the covariance matrix and reconstruct the Gaussian Wigner function of the four quadrature components
\cite{Braunstein2005}.

\begin{figure}[b]
\centering
\includegraphics[scale=.25]{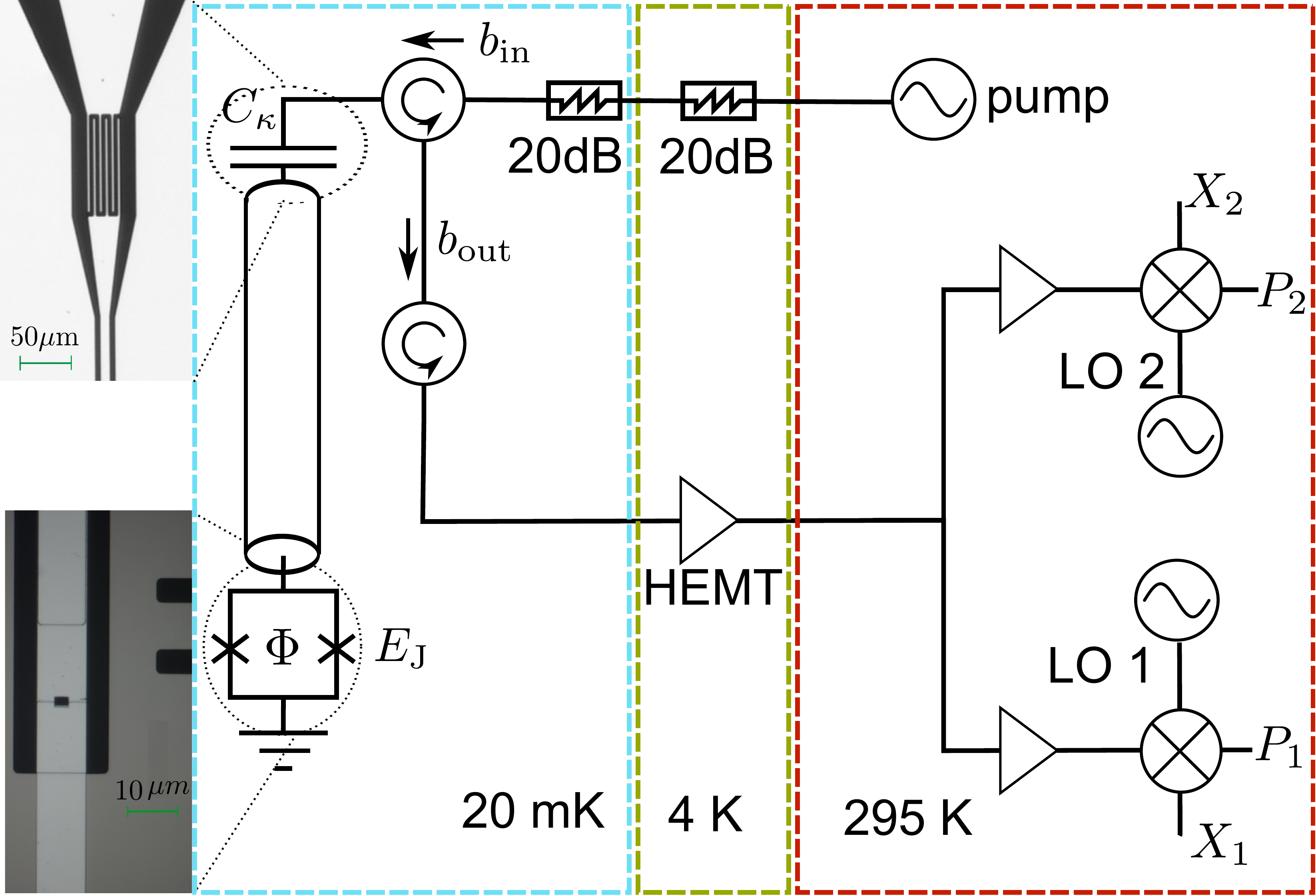}
\caption{
Schematic of the experimental setup. The parametric amplifier is realized as a pumped $\lambda/4$ resonator with a SQUID at the end. The input and and output modes are separated using a circulator. The output signal is amplified by a cold HEMT amplifier at the 4K stage and split into two channels at room temperature. Both channels have their individual local oscillator allowing for a simultaneous detection of two distinct modes in frequency space.
}
\label{fig:schematic}
\end{figure}

\begin{figure}[t]
\centering
\includegraphics[scale=.55]{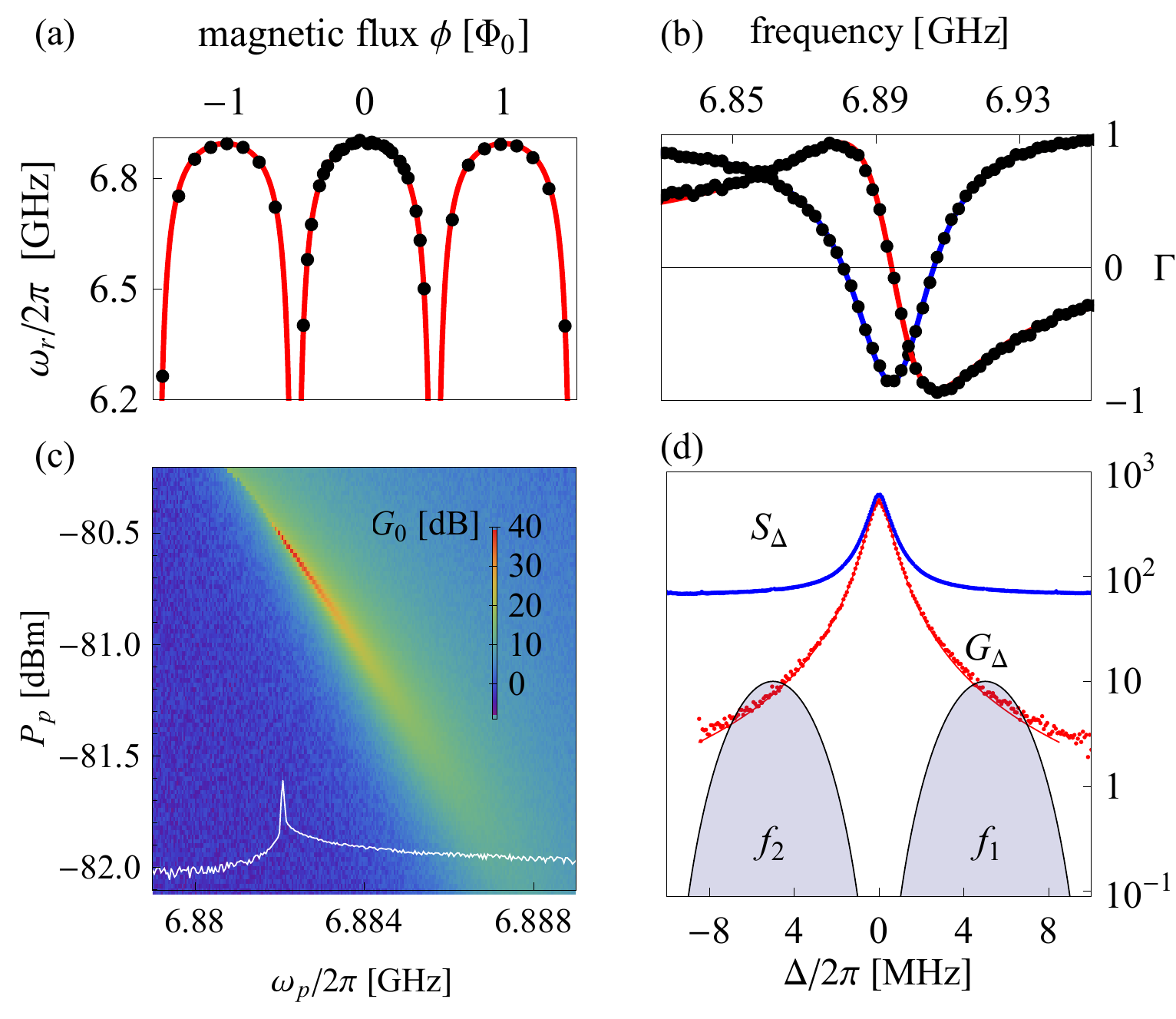}
\caption{
(a)~ Measured (dots) and calculated (solid line) resonance frequency $\omega_{\text{r}}/2\pi$ \textsl{vs.}~magnetic flux $\phi$.
(b)~Calculated real (blue line) and imaginary (red line) part of the reflection coefficient $\Gamma$ in the linear regime for magnetic flux $\phi = 0$ and measured data (dots).
(c)~Measured maximum gain $G_{0}$ as a function of pump frequency $\omega_{p}/2\pi$ and pump power $P_p$. A horizontal cut through the data
where $P_p \approx P_{\rm p,crit}$ is shown in white.
(d)~Measured gain $G_{\Delta}$ (red) and power spectral density $S_\Delta$ (blue) for a fixed pump tone as a function of signal detuning $\Delta$. The frequency filter functions $f_1(\Delta)$ and $f_2(\Delta)$, which define the modes $b_1$ and $b_2$, are shown on a logarithmic scale (arb. units) as the lines enclosing the shaded areas.
}
\label{fig:summary}
\end{figure}

Our parametric amplifier [Fig.~\ref{fig:schematic}] is realized as a $\lambda/4$ transmission line resonator terminated by a superconducting quantum interference device (SQUID) \cite{Sandberg2008}.  The fundamental mode of the resonator is in good approximation described as a nonlinear oscillator with an effective Hamiltonian
\begin{equation}
H = \hbar \omega_{\text{r}} a^\dagger a + \hbar \frac{K}{2}  a^\dagger  a^\dagger a a ,
\label{eq:Hamiltonian}
\end{equation}
where the non-linearity is provided by the SQUID. The Kerr constant $K$ as well as the resonator frequency $\omega_{\text{r}}/2\pi$ are functions of the effective Josephson energy $E_{\rm J}$ of the SQUID \cite{Wallquist2006}. Thus, by changing the flux bias $\phi$ through the SQUID loop the resonator frequency and the Kerr constant can be tuned. A measurement of the flux dependent resonance frequency allows to extract both the Josephson energy and the Kerr constant. Fitting the data shown in Fig.~\ref{fig:summary}(a) to a model described in Ref.~\cite{Wallquist2006} leads to the approximate values $E_{\rm{J,max}}/h \approx 6.1\,\rm{THz}$   and $K/\omega_{\text{r,max}} \approx -2.8\times10^{-7}$ at the maximum resonance frequency where $\omega_{\text{r,max}}/2 \pi \approx 6.9\,\rm{GHz}$. To further characterize the resonator we have measured the reflection coefficient $\Gamma$ in the linear regime [Fig.~\ref{fig:summary}(b)]. From the real and imaginary part of $\Gamma$ we can extract the coupling of the resonator to the transmission line $\kappa/2\pi \approx 25\,\rm{MHz}$ which dominates over the internal loss $\gamma_{\rm i}/2\pi \approx 2\,\rm{MHz}$.

If a nonlinear oscillator is pumped close to its bifurcation point with a coherent pump tone at frequency $\omega_{\rm p}/2\pi$
the relation between input and output field modes $b_{\text{in}}(\Delta)$ and $b_{\text{out}}(\Delta)$ can be written as \cite{Yurke2006}
\begin{eqnarray}
b_{\text{out}}(\Delta) &=& A_{\Delta} b_{\text{in}}(\Delta) +  B_{\Delta} b_{\text{in}}^\dagger({-\Delta}) \,\,.
\label{eq:inOutRelation}
\end{eqnarray}
Here $\Delta$ is the detuning from the pump frequency $\omega_{\rm p}/2\pi$, where $\Delta > 0$ corresponds to frequency components at the upper sideband of the pump whereas $\Delta < 0$ corresponds to the lower sideband.
The frequency dependent coefficients $A_{\Delta}$ and $B_{\Delta}$ fulfill the relation $|A_{\Delta}|^2 - |B_{\Delta}|^2 = 1$.  Thus Eq.~(\ref{eq:inOutRelation}) describes a minimal form quantum linear amplifier with gain $G_{\Delta}=|A_{\Delta}|^2$ \cite{Caves1982}. When a signal is applied at one sideband the frequency components of the other sideband are usually called idler modes. The device acts as a  phase-insensitive amplifier when only the signal modes are analyzed at its output  \cite{Nha2010}.

We have characterized our device as such an amplifier. To measure the maximum gain $G_{0}$ as a function of $\omega_{\rm p}/2\pi$ and pump power $P_{\rm p}$ we apply an additional weak coherent tone with a small detuning of  $\Delta/2\pi = 2.5\,\rm{kHz}$ to the input port. We measure the reflected amplitude at this frequency and compare the result with the one that we obtain when the pump tone is turned off. The absolute square of this ratio is the gain $G_0$, see Fig.~\ref{fig:summary}(c). We identify a critical point at $\omega_{\rm{p,crit}}/2\pi \approx 6.882\,\rm{GHz}$ and $P_{{\rm{p,crit}}} \approx -80.6\,\rm{dBm}$, where $G_{0}$ takes its largest value~\cite{Vijay2009}. For pump powers below $P_{{\rm{p,crit}}}$ we are in the stable amplifier regime. A decrease in pump power leads to smaller gain but to larger amplifier bandwidth $B$. The gain-bandwidth product remains constant according to the relation $\sqrt{G_0} B \propto \kappa$ \cite{Yurke2006}, which we have verified experimentally.

For the following measurements we have fixed the coherent pump at $\omega_{\rm p}/2\pi = 6.8834$ GHz and $P_{\rm p} \approx -80.8$ dBm. The signal frequency dependence of the gain $G_{\Delta}$ for this pump tone is shown in Fig.~\ref{fig:summary}(d) together with a measurement of the power spectral density $S_{\Delta}$. The power spectral density can be decomposed into two contributions
\begin{equation}
S_{\Delta} \delta(\Delta-\Delta')=\langle{b}^\dagger_{\rm out}(\Delta){b}_{\rm out}(\Delta')\rangle + N_{\text{noise}}\delta(\Delta-\Delta')\,\,.
\end{equation}
The first term on the right hand side stands for the noise at the output of the parametric amplifier. The second term describes the white noise contribution of the HEMT amplifier. Since the power spectral density at the parametric amplifier output can be interpreted as the amplified vacuum noise, the curves for $G_{\Delta}$ and $S_{\Delta}$ are almost identical up to the constant offset $N_{\text{noise}}= 69$. In the large gain limit both curves are well described by a Lorentzian function  \cite{Yurke2006}.

Up to now we have characterized our device as a phase insensitive amplifier.
If the input of the parametric amplifier is in the vacuum state, which is approximately realized with a mean thermal photon number below $\bar{n} \approx 0.05 $ as verified in independent experiments \cite{Fink2010}, the output at a single frequency component is the amplified vacuum noise. However, if one analyzes the signal and idler modes individually the parametric amplifier transforms any input state into a state where the signal and idler modes show squeezing correlations. We observe these correlations for a vacuum input state by pumping the resonator with the same coherent tone as before and recording the quadrature amplitudes in both modes individually.

The resonator output is further amplified with a HEMT amplifier and split into two channels. To separate signal and idler frequency components from each other [Fig.~\ref{fig:schematic}] we effectively down-convert the microwaves in both channels by mixing them with local oscillator tones at frequencies $\omega_{\text{LO1}}/2\pi$ and $\omega_{\text{LO2}}/2\pi$ set 5 MHz above and below the pump frequency, respectively. The voltages in both channels are digitized every $10 \,\rm{ns}$ with an analog to digital converter (ADC).  Using a field programmable gate array (FPGA) the data is digitally filtered with a symmetric filter function $f({\Delta})$. As a result
the two detection channels capture information about frequency components in the windows $f_1(\Delta) = f(\Delta - 2\pi \times  5 \rm{MHz})$ and $f_2(\Delta) = f(\Delta + 2\pi \times 5 \rm{MHz})$, respectively [Fig.~\ref{fig:summary}(d)]. The filter is designed such that both $f_1(0)=f_2(0)=0$, filtering out the coherent pump tone.

As a result, the complex numbers $S_1$ and $S_2$ that we extract after the digital data processing correspond to measurement results of the complex valued operators \cite{daSilva2010}
\begin{equation}
\hat{S}_{1,2} \equiv b_{1,2} +  h_{1,2}^\dagger \equiv \hat{X}_{1,2} + i \hat{P}_{1,2}
\label{eq:S}
\end{equation}
where
\begin{equation}
b_{1,2} = \int_{-\infty}^\infty \text{d}\Delta f_{1,2}(\Delta)  b_{\text{out}}(\Delta)\,\,.
\end{equation}
$b_{1}$ and $ b_{2}$ describe a pair of signal and idler modes at the parametric amplifier output. The modes $h_1$ and $h_2$ are in thermal states with approximately the same mean photon number $N_{\rm{noise}}$ due to the noise added by the detection chain.

Taking the filter function into account we get an operator relation similar to Eq.~(\ref{eq:inOutRelation}) for $b_1$, $b_2$ and their counterparts at the input of the parametric amplifier. Such an operator identity is equivalent to the two-mode squeezing transformation described by the unitary operator $U(r) = \text{exp}[r(b_1 b_2- b_1^\dagger b_2^\dagger )]$ \cite{Gerry2005}. The relative phase between the two local oscillators has been chosen such that the squeezing parameter $r$ is real and given by
\begin{equation}
\text{cosh}^2(r) =  \int_{-\infty}^\infty \text{d}\Delta |f_{1}(\Delta)|^2  G_{\Delta}\,\,.
\end{equation}
From our gain measurement data and the designed filter function we obtain the numerical value $r \approx 1.75$. A parametric amplifier device thus ideally transforms a vacuum input state into a two-mode squeezed vacuum state $U(r)|00\rangle$
\begin{figure}[t]
\centering
\includegraphics[scale=2]{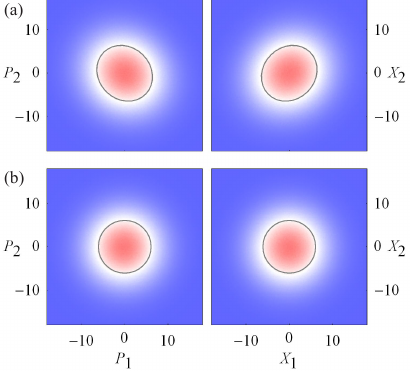}
\caption{
Measured histograms for the quadrature pairs $\{X_1,X_2 \}$ and $\{P_1,P_2\}$ with (a) pump tone on and (b) off. The gray ellipses indicate lines of constant probability density.
}
\label{fig:histograms}
\end{figure}
which shows strong correlations among the 4 quadrature components ${\hat{\xi}}_i \in \{\hat{x}_1, \hat{p}_1, \hat{x}_2, \hat{x}_2\}$ of the two modes. Those are defined by $b_{1,2} = \hat{x}_{1,2} + i \hat{p}_{1,2}$. The relative "position" $\hat{x}_1 - \hat{x}_2$ and the total "momentum" $\hat{p}_1 + \hat{p}_2$ are squeezed according to $\langle( \hat{x}_1 - \hat{x}_2 )^2\rangle = \langle( \hat{p}_1 + \hat{p}_2 )^2\rangle = e^{-2r}/2$, whereas each component itself is amplified $\langle \hat{\xi}_i^2 \rangle  = \text{cosh}(2r)$/4.
Since the two-mode squeezed state belongs to the class of Gaussian states, its Wigner function can be written as a multivariate normal distribution \cite{Braunstein2005}
\begin{equation}
W(\alpha) = \frac{1}{4 \pi^2 \sqrt{\text{det} {\bf V}}}\text{exp} \Big\{ -\frac{1}{2}\alpha {\bf V} ^{-1} \alpha^T 
 \Big\}
\label{eq:Wigner}
\end{equation}
with the vector of quadrature components $\alpha = ({x}_1,{p}_1,{x}_2,{p}_2)$ and the quadrature covariance matrix $\bf{V}$ with elements $V_{i,j} = \langle \hat{\xi}_i \hat{\xi}_j + \hat{\xi}_j \hat{\xi}_i\rangle/2$ \cite{Braunstein2005}. The four-dimensional Wigner function of the two-mode phase space distribution is thus fully determined by the $4\times4$ covariance matrix $\bf{V}$. The elements of this matrix describe the joint statistics of the amplitude fluctuations of the two modes.

To determine these elements we detect the four quadrature components as explained above and store the results in two-dimensional histograms for the six possible pairs $\{X_1,P_1\}$, $\{X_2,P_2\}$,  $\{X_1,P_2\}$, $\{X_2,P_1\}$, $\{X_1,X_2\}$ and $\{P_1,P_2\}$. For each pair we first acquire a reference histogram with the pump turned off, which characterizes the quadrature distribution of the effective noise modes $h_1$ and $h_2$ and a second histogram with the pump turned on [Fig.~\ref{fig:histograms}]. The single mode histograms $\{X_1,P_1\}$, $\{X_2,P_2\}$ (data not shown) remain circular symmetric whereas the cross-histograms $\{X_1,X_2\}$ and $\{P_1,P_2\}$ are squeezed along the diagonal axis. Because of the relatively large added amplifier noise the differences between measurements with pump on or off are small but nevertheless directly visible in the bare histograms. To scale $X$ and $P$ axes of the histograms we have used the information about the added amplifier noise extracted from the measurement of the power spectral density $S_{\Delta}$. The axes are scaled such that the histogram variances in the "pump off" measurements are equal to the noise offset $N_{\rm{noise}}$. Furthermore, we have corrected for the small difference of gain in the two detection channels.

\begin{figure}[t]
\centering
\includegraphics[scale=1.7]{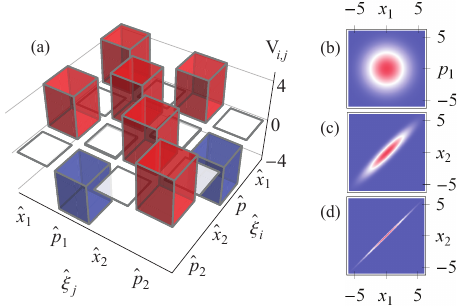}
\caption{
(a)~Measured quadrature covariance matrix $\bf{V}$ of the two-mode correlated state which allows to reconstruct the four-dimensional Wigner function. (b)~and (c) show projections of this function onto two-dimensional subspaces. (d)~For comparison, we show the same projection of the theoretical Wigner function of an ideal two-mode squeezed vacuum state.
}
\label{fig:barChart}
\end{figure}

The measured data is further analyzed by separating the contributions of modes $h_1$ and $h_2$ from those of modes $b_1$ and $b_2$. We calculate all possible expectation values $\langle \hat{X}_i \hat{X}_j \rangle$, $\langle \hat{X}_i \hat{P}_j \rangle$ from the 12 measured histograms. Taking into account that modes $h_i$ and $b_j$ are uncorrelated and using Eq.~(\ref{eq:S}), we can determine all second order expectation values $ \langle \hat{\xi}_i \hat{\xi}_j\rangle$ \cite{Eichler2010,Menzel2010}, which are summarized in the covariance matrix shown in Fig.~\ref{fig:barChart}(a). The diagonal elements in the matrix express the amplified individual quadrature fluctuations in both modes, whereas the non-vanishing off-diagonal elements describe the squeezing correlations between the two modes.  If we fit our measurement results to the ideal values for which the diagonal terms are ${\rm cosh}(2 r)/4$ and the four non-vanishing off diagonal terms $\pm {\rm sinh}(2 r)/4$, we find $r \approx 1.78$, which is close to the value that we have calculated from the independent gain measurement data and the designed filter function ($r\approx1.75$).

We note that the measured diagonal elements are approximately 3\% larger and the off-diagonal elements the same amount smaller than the ideally expected ones. The output of the parametric amplifier in modes $b_1$ and $b_2$ is thus not perfectly described by the probability distribution of a pure two-mode squeezed vacuum state but by its convolution with a thermal distribution. This thermal noise adds to the individual quadrature fluctuations and reduces the effective squeezing. Finite internal loss is one of the potential sources of uncorrelated noise \cite{Yurke2006}.

We evaluate Eq.~({\ref{eq:Wigner}}) to reconstruct the four-dimensional Wigner function $W(\alpha)$ for the two modes. In Fig.~{\ref{fig:histograms}}(b) and (c) we show projections of $W(\alpha)$ on two-dimensional subspaces. The $\{{x}_1,{p}_1\}$-projection, which describes the individual state of mode $b_1$, is thermal. Compared to the vacuum Wigner function (not shown) it has a larger variance, indicating the amplified vacuum noise. The $\{{x}_1,{x}_2\}$-projection is squeezed along the diagonal axis, visualizing the correlations between the two modes. Due to the additional uncorrelated noise the reconstructed distribution is broader than it would be in the case of an ideal squeezed two-mode vacuum state (see Fig.~{\ref{fig:histograms}}(d)) as explained above.

In summary, we have reconstructed the full quadrature covariance matrix of a two-mode correlated state to determine its four-dimensional Gaussian Wigner function. In this context, we have fully characterized a Josephson parametric amplifier device. Our experiments show that linear photon detection allows for efficient state tomography measurements of two radiation field modes.
We have observed a squeezing parameter of $r \approx 1.78$ and believe that this  value can easily be increased by building a parametric amplifier with larger bandwidth. This should improve the ratio between amplifier and detection bandwidth which furthermore allows to use the device as a low noise amplifier in other circuit QED experiments \cite{Vijay2010}. In addition, combining parametric amplifier devices in networks with beamsplitters and  superconducting qubits could allow for future continuous variable quantum computation with propagating microwave photons \cite{Braunstein2005}.
\begin{acknowledgments}
The authors would like to acknowledge fruitful discussions with Barry Sanders. This work was supported by the European Research Council (ERC) through a Starting Grant and by ETHZ.
S.~F. acknowledges the Australian Science Foundation (FWF) for support.
\end{acknowledgments}
\bibliographystyle{apsrev}

\end{document}